\documentclass[showpacs,aps,amssymb,floatfix,prd,amsmath,preprintnumbers]{revtex4}
\RequirePackage{float}
\usepackage{epstopdf}
\usepackage{hyperref}
\usepackage{capt-of}
\usepackage{graphicx}  
\usepackage{dcolumn}   
\usepackage{float}   
\usepackage{bm}
\begin{document}

\title{\bf Gravitational Baryogenesis in Non-Minimal Coupled $f(R,T)$ Gravity}

\author{
P.K. Sahoo\footnote{ Department of Mathematics, Birla Institute of
Technology and Science-Pilani, Hyderabad Campus, Hyderabad-500078,
India,  Email:  pksahoo@hyderabad.bits-pilani.ac.in},  
Snehasish Bhattacharjee\footnote{Department of Astronomy, Osmania University, Hyderabad-500007,
India,  Email: snehasish.bhattacharjee.666@gmail.com}
}

\affiliation{ }

\begin{abstract}
Baryogenesis refers to the theoretical process that occurred in the early history of the universe producing excess of matter over antimatter. Hitherto, fundamental physics which can give rise to such events is unknown. Many theories have emerged which may be able to suffice this conundrum, although concrete understanding is obscure. Gravitational baryogenesis is one such theory introduced by H. Davoudiasl, et al., Phys. Rev. Lett. \textbf{93} (2004) 201301. In this theory, for a CP violating interaction proportional to $\partial_{\mu}R$, where R denote Ricci scalar, gravitational baryogenesis is inevitable. In this paper, we have investigated the consequences of CP violating interactions proportional to $\partial_{\mu}R$, $\partial_{\mu}T$ and $\partial_{\mu}f(R,T)$ in the framework of non-minimal f(R,T) gravity. We found that for interactions proportional to $\partial_{\mu}R$ and $\partial_{\mu}f(R,T)$, baryogenesis can be realized with suitable parameter spaces producing a net baryon asymmetry factor consistent with observations, whereas for interaction proportional to  $\partial_{\mu}T$, results in a baryon asymmetry incompatible with observations.
\end{abstract}



\maketitle
  
\section{Introduction}

Baryogenesis refers to the theoretical process that occurred in the early history of the universe producing excess of matter over antimatter. Baryogenesis still remain an open question in modern cosmology. From our current understanding it is conjectured that all particles burst into existence following same laws of physics and hence production of equal amount of matter and antimatter must lead to a zero baryon number in the universe. However from our daily experiences to current cosmological observations \cite{burles,bennet} all point towards an overwhelming dominance of matter over antimatter in the universe.

Many theories have emerged to decode this enigma by considering interactions beyond the standard model in the primordial universe \cite{nozari}. Some of these theories are Affleck-Dine Baryogenesis \cite{dine,d2,d3}, GUT Baryogenesis \cite{gut}, Spontaneous Baryogenesis \cite{spontaneous,s2,s3}, Electroweak Baryogenesis \cite{electroweak, e2}, Thermal Baryogenesis and Black hole evaporation Baryogenesis \cite{black}.

In 1967, A. Sakharov published three main conditions required to generate baryon asymmetry \cite{sakharov}. These conditions are: (i) violation of net baryon number, (ii) violation of Charge ($C$) and Charge-parity ($CP$) symmetry and (iii) interactions occurring outside of thermal equilibrium.

Many authors have reported baryon asymmetry without satisfying all the Sacharov's conditions. Davoudiasl et al., \cite{dav} reported baryon asymmetry in an expanding universe by maintaining thermal equilibrium but violating ($CP$) symmetry while Kaplan \cite{kaplan} studied baryon asymmetry in thermal equilibrium by violating $CPT$ symmetry.

This mysterious and yet observationally verified concept of Baryogenesis have been studied by many authors in the framework of modified gravity. Gravitational baryogenesis have been studied by Lambiase \cite{lambiase} and Ramos \cite{ramos} in $f(R)$ gravity,  Odintsov \cite{odin} in Gauss-Bonnet gravity, Oikonomou \cite{oiko} using $f(T)$ gravity, Nozari \cite{nozari}, Baffou \cite{baffou} in minimal $f(R,T)$ gravity and Bento \cite{bento} in Gauss-Bonnet braneworld cosmology.

Previous studies in the context of gravitational baryogenesis were mainly concentrated for minimal modified gravity models where matter is weakly coupled with gravity. Here, we try to fill the gap by reporting gravitational baryogenesis for non-minimal $f(R,T)$ gravity introduced in the literature by Harko et al. \cite{harko}. The cosmological viability of non-minimal $f(R,T)$ gravity coupled with an HEL have been studied in \cite{sahoo}. At the end we also compare our model predictions with observational constraint on baryon asymmetry.

The paper is organized as follows: in Section II we present an overview of $f(R,T)$ gravity. In Section III, the gravitational baryogenesis in $f(R,T)$ gravity model are investigated by assuming a hybrid scale factor. Section IV contains a gravitational baryogenesis interaction where baryon current is coupled with trace of energy momentum tensor. A more complete and generalized version of gravitational baryogenesis is presented in Section V. Finally,  Section VI contains the conclusion and summary of the present work.

\section{Overview of $f(R,T)$ gravity}

The late time cosmic acceleration of the universe has been explored and explained by many alternate models of classical or quantum gravity \cite{nozari}. One of the most interesting theories of modified gravity is the $f(R,T)$ gravity. This theory is built on the coupling between matter and geometry. $f(R,T)$ theory can distinguish between diverse gravitational models due to its fascinating features and consistency with observations. $f(R,T)$ gravity models can explain the transition from matter dominated phase to the late dark energy dominated phase \cite{houndjo}. The gravitational Lagrangian in $f(R,T)$ gravity is a generic function of the Ricci scalar curvature $R$ and the trace of  stress-energy-momentum tensor $T$ \cite{nozari}. The action in $f(R,T)$ gravity is given by 
\begin{equation}\label{1}
S=\frac{1}{2\kappa^{2}} \int \sqrt{-g}\left[f(R,T) + \mathcal{L}_{m}\right] d^{4}x
\end{equation} 
where $\mathcal{L}_{m}$ represents matter Lagrangian.\\
Using $\mathcal{L}_{m}$ we write stress-energy-momentum tensor of matter fields as 
\begin{equation}\label{2}
T_{\mu \nu} = \frac{-2}{\sqrt{-g}}\frac{\delta (\sqrt{-g} \mathcal{L}_{m} )}{\delta g^{\mu \nu}}
\end{equation}
By varying the action (\ref{1}) with respect to the metric yields \begin{equation}\label{3}
f^{1}_{,R}(R,T)R_{\mu \nu} + \Psi_{\mu \nu}f^{1}_{,R}(R,T)-\frac{1}{2}g_{\mu\nu}f(R,T) =\kappa^{2}T_{\mu \nu}-f^{1}_{,T}(R,T)(T_{\mu \nu} + \Theta_{\mu \nu})
\end{equation}
where 
\begin{equation}\label{4}
\Psi_{\mu \nu}= g_{\mu \nu}\square-\nabla_{\mu} \nabla_{\nu}
\end{equation}
\begin{equation}\label{5}
\Theta_{\mu \nu}\equiv g^{\alpha \beta}\frac{\delta T_{\alpha \beta}}{\delta g^{\mu \nu}}
\end{equation}
and $f^{i}_{,X}\equiv \frac{d^{i}f}{d X^{i}}$. The field equations (\ref{3}) reduces to standard GR form when $f(R,T)\equiv R$.\\
Upon contracting equation (\ref{3}) with inverse metric $g^{\mu \nu}$, one obtain the trace of the field equations as \begin{equation}\label{6}
f^{1}_{,R}(R,T)R -2f(R,T) + 3\square f^{1}_{,R}(R,T)= \kappa^{2}T- f^{1}_{,T}(R,T) (T+\Theta) 
\end{equation}
We now consider a spatially flat FLRW metric as \begin{equation}\label{7}
ds^{2}=dt^{2}-a^{2}(t)[dx^{2}+dy^{2}+dz^{2}]
\end{equation}
where $a(t)$ represents the scale factor. We assume matter content of the universe to be a perfect fluid and hence matter Lagrangian density can assumed $\mathcal{L}_{m}=-p$. Applying this to equations (\ref{3}) and (\ref{6}) we obtain \begin{equation}\label{8}
3H^{2}=\frac{1}{f^{1}_{,R}(R,T)}\left[pf^{1}_{,T}(R,T) -3 \dot{R} H  f^{2}_{,R}(R,T) + \frac{1}{2}\left( f(R,T)-Rf^{1}_{,R}(R,T)\right)  \right]+\frac{\kappa^{2}+f^{1}_{,T}(R,T)}{f^{1}_{,R}(R,T)} \rho
\end{equation}
\begin{multline}\label{9} 
-3H^{2} -2\dot{H}=  \frac{1}{f^{1}_{,R}(R,T)} \left[ \ddot{R} f^{2}_{,R}(R,T) +\dot{R}^{2} f^{3}_{,R}(R,T) -\frac{1}{2}\left( f(R,T)-R f^{1}_{,R}(R,T)\right)  -pf^{1}_{,T}(R,T) + 2H \dot{R}f^{2}_{,R}(R,T) \right] \\ +\frac{\kappa^{2}+f^{1}_{,T}(R,T)}{f^{1}_{,R}(R,T)}p 
\end{multline}
where overhead dot represent time derivative, $H$ is the Hubble parameter, $\rho$ represents density and $p$ represents pressure such that $T=\rho - 3p$.

\section{Gravitational Baryogenesis in $f(R,T)$ gravity}

We now show how $f(R,T)$ gravity suffices the baryogenesis problem. Baryogenesis is calculated by an important parameter known as baryon asymmetry factor $\eta_{B}$ given by \begin{equation}\label{10}
\eta_{B}=\frac{n_{B}- \tilde{n}_{B}}{s}
\end{equation}
in which $n_{B}$ and $\tilde{n}_{B}$ represents baryon and anti-baryon number respectively and $s$ denote entropy of the universe. Current Observations like CMB and Big Bang Nucleosynthesis constrained the baryon asymmetry factor to be $\eta_{B} \leq 9 \times 10^{-11}$ \cite{ramos,ade,kolb}. A mechanism that give rise to a observationally acceptable baryon asymmetry while maintaining thermal equilibrium is now investigated. The $CP$-violating interaction term reads \begin{equation}\label{11}
\frac{1}{M^{2}_{*}}\int \sqrt{-g}J^{\mu}\partial_{\mu}R d^{4}x
\end{equation}
in which $M_{*}$ represents cutoff scale of the underlying effective theory \cite{nozari}, $R$ and $g$ being Ricci scalar and determinant of the metric respectively. $J^{\mu}$ represent baryonic current. The net baryon number density ($n_{B}$) at equilibrium ($T\gg m_{B}$) is given by $6n_{B}=T^{2}\mu_{B}g_{b}$, where $\mu_{B}=\frac{\dot{R}}{M^{2}_{*}}$ is the net baryonic chemical potential and $g_{b}$ represents total number of intrnsic degrees of freedom of baryons \cite{nozari}. Hence the baryon to entropy ratio in an accelerating universe when the temperature ($T$) falls below the critical temperature $T_{D}$ is given by \cite{dav} \begin{equation}\label{12}
\frac{n_{B}}{s} \simeq \frac{-15 g_{B} \dot{R}}{4 \pi^{2} g_{*s} M^{2}_{*} T_{D}}
\end{equation}
where $g_{*s}=\frac{45 s}{2 \pi^{2} T^{3}}$ is the total number of degrees of freedom of particles contributing to the global entropy of the universe. $T_{D}$ represents critical temperature when interactions generating baryon asymmetry occur. For $\frac{n_{B}}{s}\neq 0$, $\dot{R}$ must be non-vanishing. \\
In the framework of standard GR \cite{dav} the equation relating $R$ and $T$ reads \begin{equation}\label{13}
T= (1-3\omega)\rho= \frac{-R}{8 \pi G}
\end{equation}
where $\omega=\frac{p}{\rho}$ is the equation of state (EoS) parameter. From this equation it is clear that standard GR cannot give rise to asymmetry as for a radiation dominated universe $\omega=1/3$ and hence $R$ vanishes. 

\subsection{Baryogenesis in $f(R,T)= R+\zeta RT$ gravity model}

\begin{figure}[H]
\center
\includegraphics[width=10cm]{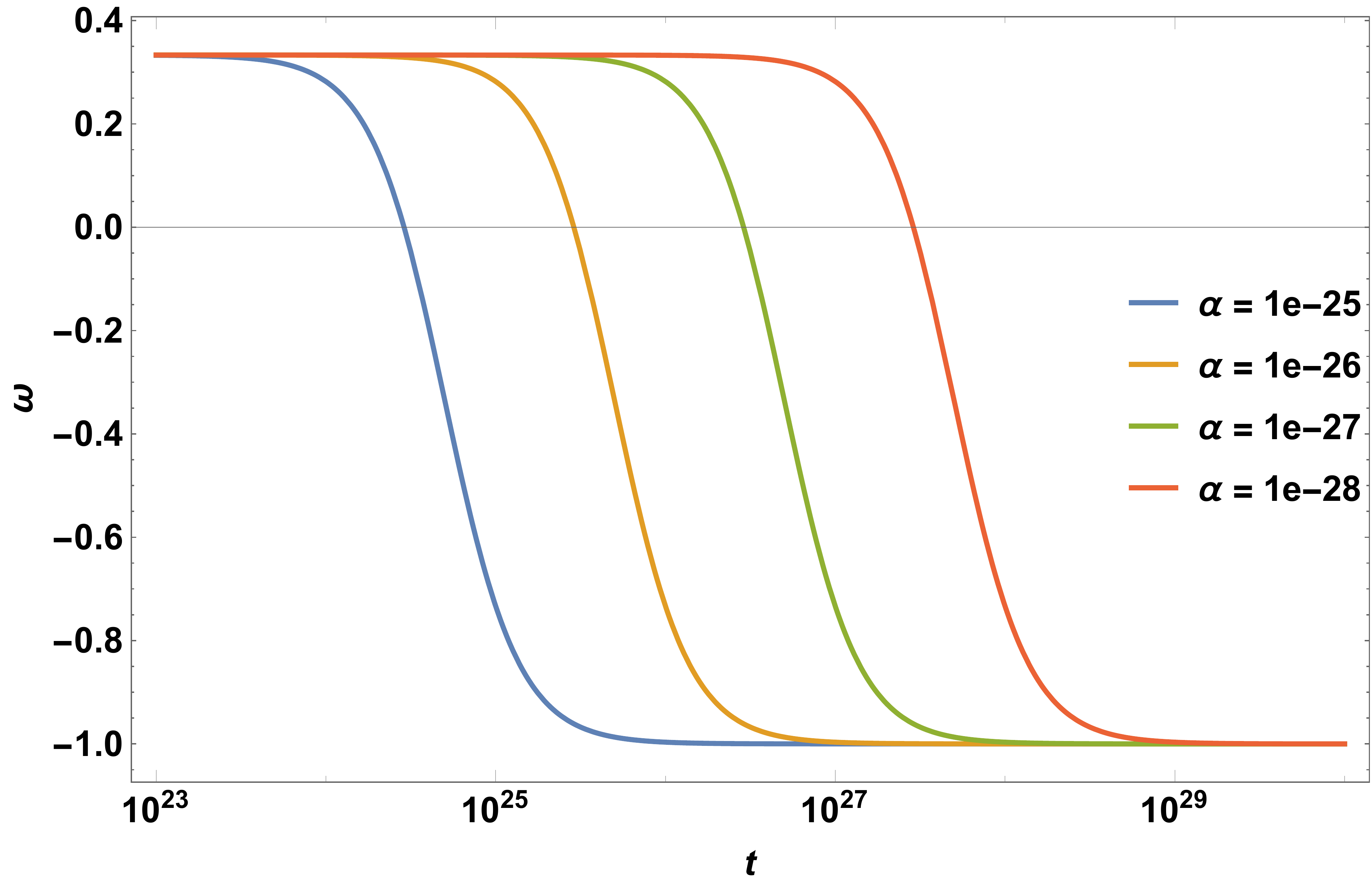}
\caption{EoS parameter ($\omega$) vs time ($t$) with different  observationally acceptable values of $\alpha$ and for $\beta = 1/2$. }
\label{f1}
\end{figure}

The simplest non-minimal matter geometry coupled $f(R,T)$ gravity model where $f(R,T)= R+\zeta RT$ capable of producing a baryon asymmetry even in a radiation dominated universe is now presented.
Substituting $f(R,T)= R+\zeta RT$ in equation (\ref{8}) and (\ref{9}) the field equations in this model reads 
\begin{equation}\label{31}
3H^{2}=\frac{p\zeta R}{1+ \zeta T} + \frac{\kappa^{2}+\zeta R}{1+ \zeta T} \rho 
\end{equation}
\begin{equation}\label{32} 
-3H^{2} -2\dot{H}=  \frac{-p\zeta R}{1 + \zeta T}  +\frac{\kappa^{2}+\zeta R}{1 + \zeta T}p 
\end{equation}
To proceed further we assume scale factor $a(t)$ of the form \cite{sahoo}
 \begin{equation}\label{14}
a(t)=e^{\alpha t}t^{\beta}
\end{equation}
where $\alpha$ and $\beta$ are constants. It will be shown that observationaly acceptable values of $\alpha$ and $\beta$ motivate the power law part to dominate at earlier epoch while the exponential part starts to tyrannize at later epochs. Hubble parameter ($H$) and its time derivative ($\dot{H}$) assume the form \begin{equation}\label{15}
H=\alpha+\frac{\beta}{t}
\end{equation}
\begin{equation}\label{16}
\dot{H}=\frac{-\beta}{t^{2}}
\end{equation}
The Ricci scalar for a flat FLRW metric is given by \begin{equation}\label{19}
R=6\left( \dot{H}+2H^{2}\right) 
\end{equation} 
By using equations (\ref{15}) and (\ref{16}) time derivative of Ricci scalar ($\dot{R}$) reads \begin{equation}\label{20}
\dot{R}=\frac{-12\beta\left[ -1+2\beta + 2\alpha t\right] }{t^{3}}
\end{equation}
By substituting equations (\ref{15}), (\ref{16}) \& (\ref{19}) into the modified Friedman equations (\ref{31}) and (\ref{32}) the following expressions of density ($\rho$) and pressure ($p$) in leading orders of $t$ reads
\begin{equation}\label{17}
\rho \simeq \frac{3\left[ \alpha^{2}t^{2}+ \beta^{2}\right] }{8\pi t^{2}}
\end{equation}
\begin{equation}\label{18}
p\simeq \frac{-3\alpha^{2}t^{2} -3\beta^{2} +2\beta}{8\pi t^{2}}
\end{equation}
From equations (\ref{17}) and (\ref{18}) we can write an expression of $\omega$ as \begin{equation}
\omega \simeq \frac{2 \beta - 3 \beta^{2} - 3\alpha^{2}t^{2}}{8\pi t^{2}}
\end{equation}
Time evolution of EoS parameter is shown in Figure \ref{f1}.

Temperature $T$ is related to energy density ($\rho$) as \cite{kolb} \begin{equation}\label{21}
\rho=\frac{\pi^{2}}{30}g_{*s}T^{4}
\end{equation}
By using equations (\ref{21}) and (\ref{17}) we obtain the decoupling time $t_{D}$ in terms of decoupling temperature $T_{D}$ as \begin{equation}\label{22}
t_{D}=\sqrt{\frac{-3\beta^{2} + 2\beta}{3\alpha^{2} + 8 \pi \omega\left[\frac{\pi^{2}}{30}g_{*s}T_{D}^{4} \right] }}
\end{equation}
Substituting equation (\ref{22}) into equation (\ref{20}) and using equation (\ref{12}) the baryon asymmetry factor reads \begin{equation}\label{23}
\frac{n_{B}}{s}\simeq\frac{45g_{b}\beta}{g_{*s}M_{*}^{2}\pi^{2}T_{D}}\left[ \frac{-1+2\beta+2\alpha\left(\frac{2\beta-3\beta^{2}}{3\alpha^{2}+\frac{4}{15}g_{*s}\pi^{3}T_{D}^{4}\omega} \right)^{0.5} }{\left(\frac{2\beta-3\beta^{2}}{3\alpha^{2}+\frac{4}{15}g_{*s}\pi^{3}T_{D}^{4}\omega} \right)^{1.5}}\right] 
\end{equation}
This clearly shows that baryon asymmetry factor is non-zero even in a radiation dominated universe (i.e,$\frac{n_{B}}{s}\neq0$ when $\omega=\frac{1}{3}$).

\begin{figure}[H]
\center
\includegraphics[width=10cm]{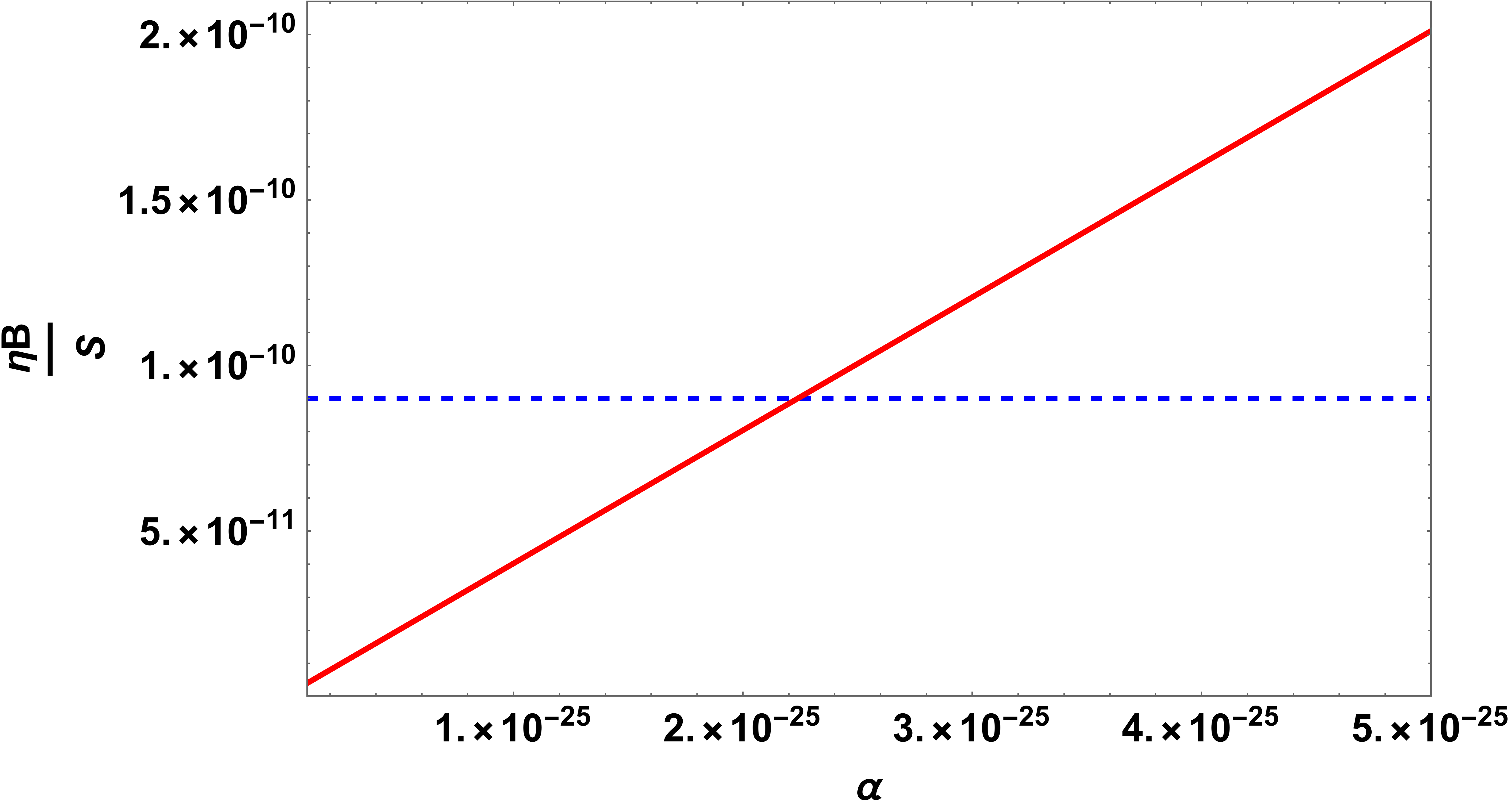}
\caption{Plot of $\frac{n_{B}}{s}$ vs $\alpha$ parameter with $\beta=0.5$. Red line indicate theoretical profile while dashed blue line represents current observational constraint. }
\label{f2}
\end{figure}

Figure: \ref{f2} shows baryon to entropy ratio as a function of $\alpha$ parameter. Following the prescription of Davoudiasl \cite{dav} we choose $M_{*}\simeq M_{planck}/ \sqrt{8\pi}$, where $M_{planck}$ represents Planck mass. The decoupling temperature $T_{D}=M_{I}\sim 2*10^{16} GeV$, where $M_{I}$ represents upper bound constraints on the tensor mode fluctuations on the inflationary scale \cite{nozari}. This particular choice of $M_{I}$ has been detected in the form of gravitational waves by LIGO. We set $g_{b}\sim \mathcal{O}(1)$ and $g_{*s}=106$. By setting $\beta=1/2$ and $\alpha=2\times10^{-25}$ we obtain baryon to entropy ratio in leading order as $n_{B}/s \sim 8.05\times10^{-11}$ which agrees well with observations. For any other values of $\beta$ other than $1/2$  and $\alpha>2.3\times10^{-25}$ we obtain physically unacceptable results. Negative $\alpha$ values results in negative baryon to entropy ratio which is again undesirable.\\
Other $f(R,T)$ gravity models of the form $f(R,T)=R+ \xi R^{v}+\chi T$ and $f(R,T)=B(1-e^{-zR})+\chi T$ also produces baryon to entropy ratio compatible with observations \cite{nozari}. However for the simplest $f(R,T)$ model where $f(R,T)=R+\chi T$ Nozari \cite{nozari} reported physically unsatisfactory results. \\
From Figure: \ref{f1} it is evident that as we decrease the value of $\alpha$ it takes more time for the universe to make a transition from being radiation dominated ($\omega = 1/3$) to being matter dominated ($\omega = 0$) and finally to the currently observed mysteriously accelerated phase ($\omega = -1$). However we must restrict the value of $\alpha$ to be $0<\alpha<2.3\times10^{-25}$ to be consistent with observational constraints on baryon to entropy ratio.

\section{Baryogenesis for Baryon current coupling with trace of Energy-momentum Tensor}

In this section gravitational baryogenesis for the model $f(R,T)= R+\zeta RT$ is investigated by considering the coupling between baryonic current $J^{\mu}$ and first order derivative of the trace of energy-momentum tensor $\partial_{\mu}T$ and finally obtain $\frac{n_{B}}{s}$ in this case. The $CP$-violation interaction in this framework is given by \cite{nozari}
\begin{equation}\label{24}
\frac{1}{M^{2}_{*}}\int \sqrt{-g}J^{\mu}\partial_{\mu}T d^{4}x
\end{equation}
For this type of interaction, the baryon to entropy ratio reads \cite{nozari}
\begin{equation}\label{25}
\frac{n_{B}}{s} \simeq \frac{-15 g_{B} \dot{T}}{4 \pi^{2} g_{*s} M^{2}_{*} T_{D}}
\end{equation}
Since $T=\rho - 3p$, by using equations (\ref{17}) and (\ref{18}) time derivative of trace of stress-energy-momentum tensor ($\dot{T}$) in leading order of $t$ reads 
\begin{equation}\label{26}
\dot{T}\simeq \frac{3\left[ 1-2\beta\right]\beta }{2\pi t^{3}}
\end{equation}
Now, after substituting equation (\ref{22}) into equation (\ref{26}) and using equation (\ref{25}) the net baryon to entropy ratio reads 
\begin{equation}\label{27}
\frac{n_{B}}{s} \simeq \frac{-45 g_{b}}{8 g_{*s} M_{*}^{2}\pi^{3}T_{D}}\frac{\left(1-2\beta \right)\beta }{\left[\frac{2\beta-3\beta^{2}}{3\alpha^{2}+\frac{4}{15}g_{*s}\pi^{3} T_{D}^{4}\omega} \right] ^{1.5}}
\end{equation}
which is non-vanishing for $\omega=1/3$. We choose as before, $M_{*}\simeq M_{planck}/ \sqrt{8\pi}$, $T_{D}=M_{I}\sim 2*10^{16} GeV$, $g_{b}\sim \mathcal{O}(1)$ and $g_{*s}=106$. However the resultant baryon to entropy ratio is very large and hence unacceptable. For $\beta=1/2$ we get $\frac{n_{B}}{s}=0$. Other $f(R,T)$ gravity models of the form $f(R,T)=R+\chi T$ and $f(R,T)=B(1-e^{-zR})+\chi T$ also provide unsuitable results \cite{nozari}.\\
However, a $f(R,T)$ gravity model of the form $f(R,T)=R+ \xi R^{v}+\chi T$ provide viable baryon to entropy ratio in this type of baryogenesis interaction \cite{nozari}.

\section{Generalized gravitational baryogenesis}

A more complete and general baryogenesis interaction term is now investigated which is given by \cite{nozari}
\begin{equation}\label{28}
\frac{1}{M^{2}_{*}}\int \sqrt{-g}J^{\mu}\partial_{\mu}f(R,T) d^{4}x
\end{equation}
The baryon to entropy ratio is given by \cite{nozari}
\begin{equation}\label{29}
\frac{n_{B}}{s} \simeq \frac{-15 g_{b} (\dot{T}f^{1}_{,T}+\dot{R}f^{1}_{,R})}{4 \pi^{2} g_{*s} M^{2}_{*} T_{D}}
\end{equation}
For the model $f(R,T)=R+\zeta RT$ we obtain $\dot{T}f^{1}_{,T}+\dot{R}f^{1}_{,R}= \dot{R}\left( 1+\zeta T\right)+ \zeta \dot{T}  R $. Substituting all the values into equation  (\ref{29}) we obtain baryon to entropy ratio as 
\begin{multline}\label{30}
\frac{n_{B}}{s} \simeq \frac{-15 g_{B}}{4 \pi^{2} g_{*s} M^{2}_{*} T_{D}}\left[ -12\beta \left( \frac{-1+2\beta+2\alpha t_{D}^{2} }{ t_{D}^{3}}\right) \left( 1+ \zeta\left( \frac{3\left( -\beta + 2\beta^{2} +2 \alpha^{2} t_{D}^{2}  \right) }{4 \pi  t_{D}^{2}  }\right) \right) \right]\\-\frac{15 g_{B}\zeta}{4 \pi^{2} g_{*s} M^{2}_{*} T_{D}}\left[ \left(\frac{3}{2\pi}\frac{\left(1-2\beta \right)\beta }{ t_{D}^{3/2}} \right)\left( \frac{-6\beta}{t_{D}^{2}} + 12\left(\alpha + \frac{\beta}{t_{D}^{2}}  \right) ^{2}\right)  \right]  
\end{multline}
where $t_{D}$ is given by equation (\ref{22}).
By setting $\beta=1/2$, the second term in equation (\ref{30})  vanishes. Choosing $T_{D}$, $g_{b}$, $g_{*s}$ as before and $\zeta=1$ and $\alpha=2\times10^{-25}$ we obtain baryon to entropy ratio in leading order as $n_{B}/s\simeq  8.03\times10^{-11}$ which is close to the observational value. To obtain acceptable results $\beta$ must be constrained to $1/2$. Other $\beta$ values generate unreasonable results. Similar acceptable baryon to entropy values were obtained by Nozari \cite{nozari} for their $f(R,T)$ model $f(R,T)=R+ \xi R^{v}+\chi T$ with other $f(R,T)$ models produced unacceptable results. Additionally $f(R,T)$ models of the form $f(R,T)=R+\phi T + \varphi T^{2}$ and $f(R,T)= R + \varsigma R^{2}+ \tau T$ also lead to physically acceptable baryon to entropy ratios \cite{baffou}.

\section{Conclusion}

In this work we studied Gravitational baryogenesis by considering the simplest non-minimal matter-geometry coupled $f(R,T)$ gravity model where $f(R,T)= R+\zeta RT$. We have shown that this $f(R,T)$ gravity model is efficient in describing the observed baryon to entropy ratio. For the first type of baryogenesis interaction involving derivative of Ricci scalar ($\partial_{\mu}R$) our model yielded theoretical value in leading order as $n_{B}/s \sim 8.05\times10^{-11}$ which is in excellent agreement with observational value of $n_{B}/s \sim 9\times10^{-11}$. Other $f(R,T)$ gravity models of the form $f(R,T)=R+ \xi R^{v}+\chi T$ and $f(R,T)=B(1-e^{-zR})+\chi T$ also produced baryon to entropy ratio compatible with observations \cite{nozari}.\\
We then consider another type of baryogenesis interaction involving coupling between baryonic current and derivative of energy-momentum tensor $\partial_{\mu}T$. Our model failed to produce any satisfactory baryon to entropy ratio in this set up. However as reported by \cite{nozari}, $f(R,T)$ gravity model of the form $f(R,T)=R+ \xi R^{v}+\chi T$ provide viable baryon to entropy ratio in this type of baryogenesis interaction.\\
We finally conclude our work by studying a more complete and generalized baryogenesis interaction which is proportional to $\partial_{\mu} f(R,T)$. In this type of interaction our model produced theoretical value in leading order as $n_{B}/s\simeq  8.03\times10^{-11}$ which is close to the observational value and also to the value obtained for the first case. Other $f(R,T)$ models of the form $f(R,T)=R+\phi T + \varphi T^{2}$, $f(R,T)= R + \varsigma R^{2}+ \tau T$ and $f(R,T)=R+ \xi R^{v}+\chi T$ also generated good theoretical values \cite{nozari,baffou}.\\  
For this work we assumed a scale factor of the form $a(t)=e^{\alpha t}t^{\beta}$ \cite{sahoo}. Since $\alpha\sim 10^{-25}$ and $\beta=0.5$ were the appropriate values needed to obtain a viable baryon to entropy ratio we conclude that for such a scale factor the power law part dominate at early times while the exponential part preside over at later epochs which is in consistent with observations \cite{observations}.\\
We also found that a decrease in observationally acceptable value of $\alpha$ delays the transition of the universe from being
radiation dominated to being matter dominated and finally to the currently observed mysteriously accelerated phase. 
\section*{Acknowledgments}
On of the author (PKS) acknowledge DST, New Delhi, India for providing facilities through DST-FIST lab, Department of Mathematics, BITS-Pilani, Hyderabad Campus, where a part of the work was done. We are very much grateful to the honorable referee and the editor for the illuminating suggestions that have significantly improved our work in terms of research quality and presentation.

\end{document}